\documentclass[twocolumn,aps,floats,letterpaper,floatfix,groupedaddress]{revtex4-1}
\usepackage{amsmath}
\usepackage{graphicx}
\usepackage{amsfonts}
\usepackage{amssymb}
\usepackage{bm}
\usepackage{dsfont}
\usepackage{epsfig,float,afterpage}
\usepackage[colorlinks,linkcolor=blue,citecolor=blue,urlcolor=blue]{hyperref}
\usepackage[usenames,dvipsnames]{color}

\newcommand{\beq}{\begin{equation}}
\newcommand{\eeq}{\end{equation}}
\newcommand{\bes}{\begin{subequations}}
\newcommand{\ees}{\end{subequations}}
\newcommand{\bea}{\begin{eqnarray}}
\newcommand{\eea}{\end{eqnarray}}
\newcommand{\ba}{\begin{array}}
\newcommand{\ea}{\end{array}}
\newcommand{\beqn}{\begin{eqnarray*}}
\newcommand{\eeqn}{\end{eqnarray*}}

\newcommand{\f}[2]{\frac{#1}{#2}}

\newcommand{\la}{\langle}
\newcommand{\ra}{\rangle}

\newcommand{\dg}{\dagger}

\def\one{1\hskip -1mm{\rm l}}

\def\nn{\nonumber}

\begin{document}
\title{Random Matrix Spectral Form Factor in Kicked Interacting Fermionic Chains
}
\author{Dibyendu Roy$^1$ and Toma$\check{\rm z}$ Prosen$^2$}
\affiliation{$^1$Raman Research Institute, Bangalore 560080, India}
\affiliation{$^2$Physics Department, Faculty of Mathematics and Physics, University of Ljubljana, Jadranska 19, SI-1000 Ljubljana, Slovenia} 

\begin{abstract}
We study quantum chaos and spectral correlations in periodically driven (Floquet) fermionic chains with long-range two-particle interactions, in the presence and absence of particle number conservation ($U(1)$) symmetry. We analytically show that the spectral form factor precisely follows the prediction of random matrix theory in the regime of long chains, and for timescales that exceed the so-called Thouless/Ehrenfest time which scales with the size $L$ as ${\cal O}(L^2)$, or ${\cal O}(L^0)$, in the presence, or absence of $U(1)$ symmetry, respectively.
Using random phase assumption which essentially requires long-range nature of interaction, we demonstrate that the Thouless time scaling is equivalent to the behavior of the spectral gap of a classical Markov chain, which is in the continuous-time (Trotter) limit generated, respectively, by a gapless $XXX$, or gapped $XXZ$, spin-1/2 chain Hamiltonian.
\end{abstract}

\maketitle
Symmetry and the related invariance of physical laws under transformations are important concepts of physics. The study of symmetry in quantum physics was primarily introduced by Wigner \cite{Wigner1959}, who also famously formulated random matrix theory (RMT) for a statistical description of fluctuations in positions of compound nuclei resonances \cite{Wigner1955}. The concepts of symmetry are closely intertwined in the development of RMT. The essential idea of the RMT is that the universal fluctuations in complicated spectra of any compound nucleus consisting of many strongly interacting nucleons can be described in terms of statistical ensembles of eigenvalues of very large random matrices with independent, identically distributed ({\em iid}) entries. The properties of input random matrices are determined by the general features of the underlying generic Hamiltonians of the nucleus, such as the hermiticity, the time-reversal symmetry, and other unitary symmetries \cite{Mehta2004, Fyodorov2011}. Following these ideas of Wigner, Dyson proposed three major symmetry classes of random matrices (namely, orthogonal, unitary, and symplectic) based on the symmetry classification of Hamiltonians \cite{Dyson1962}.

The RMT also plays a very special role in the advances of quantum chaos where a statistical similarity was observed/conjectured between the RMT spectra and highly excited energy levels of generic quantum systems whose corresponding classical dynamics are fully chaotic \cite{McDonaldPRL1979,Casati1980,Berry1981,BohigasPRL1984}. The fluctuations in spectral density (spectral fluctuations) of energy or quasienergy spectra are often used as the main signatures of quantum chaos and the appropriate RMT type is determined solely by the symmetry of 
underlying dynamical systems \cite{Haake2001}. The simplest measure of spectral fluctuations is the spectral form factor (SFF) $K(t)$, i.e., the Fourier transform of the spectral pair correlation function.  A partial explanation of the above observation/conjecture was given in a heuristic derivation of $K(t)$ for classically strongly chaotic systems using Berry's semiclassical periodic orbit theory \cite{Berry1977, Berry1985, Sieber2001,Sieber2002,MullerPRL2004,MullerPRE2005}. \
The only rigorous proof of the semiclassical quantum chaos conjecture has been possible for very specific class of models, namely the connected quantum graphs \cite{Weidenmueller}.
A series of recent works \cite{KosPRX2018, BertiniPRL2018, ChanPRL2018, ChanPRX2018} has further provided some new insights for low dimensional and locally interacting, nonintegrable many-body systems where local degrees of freedom, e.g., spin-1/2's, fermions, qubits, have no classical limit. Specially, these works explain the dynamical mechanism underlying the RMT descriptions of spectral properties of these systems \cite{Montambaux,Poilblanc,pineda} by going beyond the semiclassical periodic-orbit approaches.

Time-reversal and unitary symmetries are useful tools in understanding quantum dynamics as well as the RMT symmetry classification. Periodically driven many-body quantum systems with or without time-reversal invariance were investigated in the recent analytical derivation of $K(t)$. For example, while Ref.~\cite{KosPRX2018} considered a time-reversal invariant model of long-range Ising spin-1/2 chain in a periodically kicking transverse field, the  Floquet local Haar-random unitary (nearest-neighbor) quantum circuits without time-reversal invariance were explored in \cite{ChanPRL2018, ChanPRX2018}. Surprisingly, these studies of models with different time-reversal symmetry show the same (logarithmic) system-size scaling of the Thouless/Ehrenfest timescale beyond which $K(t)$ has universal RMT form. The above studies \cite{KosPRX2018, BertiniPRL2018, ChanPRL2018, ChanPRX2018} focused on models without any local conserved charges due to unitary symmetries. One can intuitively think that the presence of local conserved charges in the model will substantially alter the dynamics, especially short-time dynamics. Thus, it is expected to change the system-size scaling of Thouless timescale \cite{Gharibyan2018}. Last year, Ref.~\cite{FriedmanPRL2019} computed $K(t)$ analytically in the limit of local Hilbert space dimension $q\to \infty$ for a Floquet circuit model that has a $U(1)$ symmetry (conserved charge) encoded via spin-1/2 degrees of freedom. The model in \cite{FriedmanPRL2019} corresponds to the circular unitary ensemble (CUE) of RMT without time-reversal symmetry.

In this Letter, we make an important step forward by exploring analytically and numerically the role of $U(1)$ symmetry on $K(t)$ and the Thouless timescale in a minimal model of spinless fermions with finite local Hilbert space dimension ($q=2$) and with time-reversal symmetry. Our model belongs to the circular orthogonal ensemble (COE) of RMT. We particularly show that $K(t)$ in the Trotter regime (of the continuous-time limit) can be obtained from a Hamiltonian of an anisotropic and isotropic Heisenberg spin-1/2 chain model with periodic boundary conditions, respectively, in the absence and presence of $U(1)$ symmetry. While we find the Thouless time to be independent of the system size for a $U(1)$ symmetry breaking model, it scales quadratically with system size for a $U(1)$ symmetric model. We also provide exact numerical results for $K(t)$, which are consistent with our theoretical predictions.   

We consider a one-dimensional lattice of interacting spinless fermions with a time-periodic kicking in the nearest-neighbor coupling (hopping). The Hamiltonian of the driven system is given by
\bea
\hat{H}(t)&=&\hat{H}_0+\hat{H}_1\sum_{m \in \mathbb{Z}}\delta(t-m),\label{ham1} \\
\hat{H}_0&=& \sum_{i=1}^L\epsilon_i\hat{n}_i + \sum_{i<j} U_{ij}\hat{n}_i\hat{n}_j,
\eea
where the time is measured in units of pulse period (cycle) \footnote{Note that the roles of the `free' and `kick' Hamiltonians $H_0$ and $H_1$ can also be swapped as they enter symmetrically
in the expression of the Floquet propagator (\ref{EvU}).}. 
Here, $\hat{n}_i=\hat{c}_{i}^{\dg}\hat{c}_{i}$ is the number operator where $\hat{c}_{i}^{\dg}$ is a creation operator of a fermion at site $i$. The long-range interaction between fermions at sites $i$ and $j$ is given by $U_{ij}=U_0/|i-j|^{\alpha}$ and $\epsilon_i$ are random onsite energies which we select as Gaussian {\em iid} variables of zero mean and standard deviation $\Delta \epsilon$. The exponent is taken in the interval $1<\alpha<2$ where the interaction is neither short-ranged nor it can
be described by a mean-field. We choose the driving Hamiltonian $\hat{H}_1$ with or without a $U(1)$ symmetry which corresponds respectively to conservation or violation of a total fermion number operator $\hat{N}=\sum_{i=1}^L\hat{n}_i$. Thus, we take the following Hamiltonian:
\bea
\hat{H}_1=\sum_{i=1}^L(-J\hat{c}_i^{\dg}\hat{c}_{i+1}+\Delta\hat{c}_{i}^{\dg}\hat{c}^{\dg}_{i+1}+{\rm H.c.}),
\eea
where $J$ and $\Delta$ are respectively the strength of hopping and the amplitude of $p-$wave pairing. The absence or presence of pairing $\Delta$ generates $U(1)$ symmetric or symmetry-broken driving. We here use periodic boundary condition in real space, i.e., $\hat{c}_i\equiv\hat{c}_{i+L}$. Our above model is a long-range and disordered variant 
of the kicked $t$-$V$ model \cite{ProsenPRL1998,ProsenPRE1999} studied earlier in the context of ergodicity breaking transition in generic nonintegrable quantum many-body systems.   

For a periodically driven Floquet system as ours, the quasienergies of interest are the eigenphases of a unitary Floquet propagator $\hat{U}$ of evolution after one cycle: $\hat{U}=\mathcal{T}{\rm exp}(-i\int_0^1dt\hat{H}(t))$ where $\mathcal{T}$ is time-ordering. We then define: $\hat{U}|m\ra=e^{-i\varphi_m}|m\ra$ for $m=1,2,\dots,\mathcal{N}$ where $|m\ra$ and $\varphi_m$ are formal many-particle eigenstates and eigenphases of the Floquet system, and $\mathcal{N}$ denotes dimension of the Hilbert space. The spectral density of the eigenphases is given by
$ \rho(\varphi)=(2\pi/\mathcal{N})\sum_{m}\delta(\varphi-\varphi_m)$, with a unit mean level density $\la \rho(\varphi)\ra_{\varphi}\equiv \int_0^{2\pi}d\varphi \rho(\varphi)/(2\pi)=1$ where the average $\la \dots \ra_{\varphi}$ is carried over the full range of eigenphases. The spectral pair correlation function $R(\vartheta)=\la \rho(\varphi+\vartheta/2)\rho(\varphi-\vartheta/2)\ra_{\varphi}-\la \rho(\varphi)\ra_{\varphi}^2$ provides a measure of spectral fluctuations. 
We define and calculate its Fourier transform as
\bea
K(t)=\f{\mathcal{N}^2}{2\pi}\int_0^{2\pi}d\vartheta R(\vartheta) e^{-i\vartheta t}=|{\rm tr}\hat{U}^t|^2-\mathcal{N}^2\delta_{t,0},\label{FormF}
\eea
where ${\rm tr}\hat{U}^t=\sum_m\:e^{-i\varphi_mt}$. However, since the quantity above is not self-averaging, we further include
an average $\la \dots \ra$ over a quench disorder $\{\epsilon_i\}$ yielding the precise definition of SFF:
\bea
K(t)=\la({\rm tr}\hat{U}^t)({\rm tr}\hat{U}^{-t})\ra-\mathcal{N}^2\delta_{t,0}.\label{SFF}
\eea
The time-evolution operator $\hat{U}$ of each cycle for the system in Eq.~(\ref{ham1}) can be written as a two-step unitary Floquet propagator:
\bea
\hat{U}=\hat{V}\hat{W},\quad \hat{W}=e^{-i\hat{H}_0}~ {\rm and}~ \hat{V}=e^{-i\hat{H}_1}. \label{EvU}
\eea
It is generally difficult to derive the exact many-particle eigenstates $|m\ra$ of $\hat{U}$. Instead, we consider a basis of Fock states $|\underline{n}\ra \equiv |n_1,n_2,\dots,n_L\ra$, where $n_j\in \{0,1\}$ represents an occupation number of spinless fermion at the lattice site $j$, and $N \equiv \la\underline{n}|\hat{N}|\underline{n}\ra=\sum_{j=1}^Ln_j$. For $\Delta=0$, $[\hat{U},\hat{N}]=0$; thus $\hat{U}$  is block-diagonal in basis $|\underline{n}\ra$. Therefore, we need to consider all or some of the basis states in our derivation of $K(t)$ depending on whether $\Delta \ne 0$ or $\Delta=0$. When we consider a $U(1)$ symmetric model with total particle number conservation, we need to take only those basis states ${\cal N}={L\choose N}$, which have that particular total number $N$ of particles, e.g., $N=L/2$ at half-filling.
However, we need take all ${\cal N}=2^L$ states $|\underline{n}\ra$ in the absence of $U(1)$ symmetry in our model. These basis states $|\underline{n}\ra$ are eigenstates of $\hat{W}$, and we can write,
\bea
\hat{W}|\underline{n}\ra=e^{-i\theta_{\underline{n}}}|\underline{n}\ra,~\theta_{\underline{n}}=\sum_{i=1}^L\epsilon_in_i + \sum_{i<j} U_{ij}n_in_j.
\eea
Following Eq.~(\ref{SFF}), we now evaluate ${\rm tr}\hat{U}^t$, which is essentially the Floquet propagator for $t$ time steps, in a discrete-path-integral like fashion by inserting decomposition
of identity $\one_{\cal N}=\sum_{\underline{n}_\tau}|\underline{n}_{\tau}\ra \la\underline{n}_{\tau}|$ at different time steps $\tau=1,2,\dots,t$. The sums are restricted to an appropriate subset of
basis states in the particle-number conserving case. Thus, we write
\bea
&&{\rm tr}\hat{U}^t=\sum_{\underline{n}_1,\dots,\underline{n}_t}\la \underline{n}_1|\hat{V}\hat{W}|\underline{n}_2\ra\la \underline{n}_2|\hat{V}\hat{W}\dots|\underline{n}_t\ra\la \underline{n}_t|\hat{V}\hat{W}|\underline{n}_1\ra \nn \\
&&~~~~~~=\sum_{\underline{n}_1,\dots,\underline{n}_t}e^{-i\sum_{\tau=1}^t\theta_{\underline{n}_{\tau}}}\prod_{\tau=1}^t V_{\underline{n}_{\tau},\underline{n}_{\tau+1}},\\
&&V_{\underline{n}_{\tau},\underline{n}_{\tau+1}}=\la \underline{n}_{\tau}|\hat{V}|\underline{n}_{\tau+1}\ra,
\eea
where we use periodic boundary condition in time $\underline{n}_{t+1}\equiv \underline{n}_1$.  
We thus get the following expression of the SFF:
\bea
K(t)&=&\sum_{\underline{n}_1,\dots,\underline{n}_t}\sum_{\underline{n}'_1,\dots,\underline{n}'_t}\la e^{-i\sum_{\tau=1}^t(\theta_{\underline{n}_{\tau}}-\theta_{\underline{n}'_{\tau}})}\ra \nn\\
&\times&\prod_{\tau=1}^t V_{\underline{n}_{\tau},\underline{n}_{\tau+1}}V^*_{\underline{n}'_{\tau},\underline{n}'_{\tau+1}}.
\eea
The phases $\theta_{\underline{n}_{\tau}}$ for different many-particle basis states $\underline{n}_{\tau}$ (modulo $2\pi$) are assumed to be independent random numbers for 
$\hat{H}_0$ with random onsite energies and long-range interaction. We confirm later the validity of such random phase assumption (RPA) by direct numerical evaluation of $K(t)$. 
Within the RPA, the averaging over the disorder realizations gives:
\bea
\la e^{-i\sum_{\tau=1}^t(\theta_{\underline{n}_{\tau}}-\theta_{\underline{n}'_{\tau}})}\ra \simeq\delta_{\{\underline{n}_1,\dots,\underline{n}_t\},\{\underline{n}'_1,\dots,\underline{n}'_t\}},\label{RPA}
\eea
where $\{\underline{n}_1,\underline{n}_2,\dots,\underline{n}_t\}$ represents a lexicographically ordered string of configurations.
We assume that our system is large enough, so that at times of our interest $t \ll t_H$ where $t_H=\mathcal{N}$ is the so-called Heisenberg time.
Since the probability that two configurations repeat in time $t$ is proportional to $t/t_H$,  under condition $t\ll t_H$ all configurations $\underline{n}_{\tau}$ in the string $\{\underline{n}_1,\underline{n}_2,\dots,\underline{n}_t\}$ can be assumed different. Then, the constraint in Eq.~(\ref{RPA}) implies that there is a permutation $\pi\in S_t$ so that we can relate $\underline{n}'_{\tau}=\underline{n}_{\pi(\tau)}$. There are $t$ cyclic permutations and $t$ anti-cyclic permutations for which the matrices $V_{\underline{n}_{\tau},\underline{n}_{\tau+1}}$ in a string 
$\{\underline{n}_1,\underline{n}_2,\dots,\underline{n}_t\}$ are the same as $V_{\underline{n}'_{\tau},\underline{n}'_{\tau+1}}$ in a string $\{\underline{n}'_1,\underline{n}'_2,\dots,\underline{n}'_t\}$. One can show that the contributions to the $K(t)$ from all other permutations $\pi$ which contain at least one pair of neighbor changes are smaller in the thermodynamic limit \cite{KosPRX2018}. Thus, we can write for the leading order contributions:
\bea
K(t)&=&\sum_{\underline{n}_1,\dots,\underline{n}_t}\prod_{\tau=1}^t V_{\underline{n}_{\tau},\underline{n}_{\tau+1}}V^*_{\underline{n}_{\pi({\tau})},\underline{n}_{\pi({\tau+1})}}\nn\\
&=&2t\sum_{\underline{n}_1,\dots,\underline{n}_t}\prod_{\tau=1}^t |V_{\underline{n}_{\tau},\underline{n}_{\tau+1}}|^2=2t\:{\rm tr}\mathcal{M}^t,  \label{Mmat}
\eea
where 
\bea
\mathcal{M}_{\underline{n},\underline{n}'}=|\la \underline{n}|\hat{V}|\underline{n}'\ra|^2 = |\la \underline{n}|e^{-i H_1}|\underline{n}'\ra|^2
\label{markov}
\eea
is a  $\mathcal{N}\times \mathcal{N}$ square matrix.

In fact, $\mathcal{M}$ is a double stochastic matrix, i.e., the sums of its nonnegative elements along rows and columns are equal to $1$,
hence its largest eigenvalue is 1 as a consequence of unitarity of $\hat{V}$.  Let us write the eigenvalues of $\mathcal{M}$ as $1,\lambda_1,\lambda_2,\lambda_3,\dots$ with $1\ge |\lambda_j| \ge |\lambda_{j+1}|$. 
Then, we obtain SFF as a sum over eigenvalues $\lambda_j$
\bea
K(t)=2t\big(1+\sum_{j}\lambda_j^t\big), \label{series}
\eea
where $K(t)\simeq 2t$ is a leading order in $t/t_{\rm H}$ result of RMT/COE.
For large enough $L$, we can approximate the above expression at long time $t$, $1\ll t\ll t_H$, by truncating it after the second largest eigenvalue $\lambda_1$ of $\mathcal{M}$. The eigenvalues $\lambda_j$ may or may not depend on the dimension $\mathcal{N}$ of $\mathcal{M}$ which itself depends on $L$. Let us consider that $\lambda_1$ scales with system size $L$ as $1-1/t^*(L)$
where $t^*(L)\simeq L^\beta/D$, $D$ is a constant and $\beta$ is an exponent to be determined. Here, $t^*$ can be identified as the Thouless time by borrowing the notion from diffusive disordered conductors \cite{ThoulessPRL1977, Altshuler1986, ChanPRL2018}, or a many-body analogue of the Ehrenfest time \cite{KosPRX2018}. Thus, we can write
\bea
K(t)\simeq 2t(1+(1-1/t^*(L))^t)\simeq 2t(1+e^{-t/t^*(L)}).\quad
\eea
Our next goal is to find the $L$-dependence of $\lambda_1$ and $t^*$ in the presence and absence of $U(1)$ symmetry in our model. For this, we now consider the matrix $\mathcal{M}$ in the Trotter regime at small $J, \Delta$. Remarkably, the double stochastic matrix $\mathcal{M}$ can then be generated by a hermitian ``quantum'' Hamiltonian of the anisotropic Heisenberg (an $XXZ$ or rather an $XZZ$) model with periodic boundary condition. For $J, \Delta \to 0$, we can express the matrix $\mathcal{M}$ as 
\bea
\mathcal{M}=(1-c_x L)\one_{\cal N}+\sum_{j=1}^L\sum_{\nu} c_{\nu}\sigma^{\nu}_j\sigma^{\nu}_{j+1}+\mathcal{O}(J^4,\Delta^4), \qquad \label{map}
\eea
 where $c_x=(J^2+\Delta^2)/2,~c_{y}=c_{z}=(J^2-\Delta^2)/2$. Here, $\sigma^{\nu}_j$, $\nu\in\{x,y,z\}$, is the Pauli matrix at site $j$. The ``ground state'' of the generating Hamiltonian with an eigenvalue $1$ is a ferromagnet polarised in $x$-direction with all 
${\cal N}$ components in computational $\sigma^z_j$-basis equal and corresponds to the equilibrium state of the Markov chain. 
It matches with our previous claim of largest eigenvalue $\lambda_0=1$ for the double stochastic matrix $\mathcal{M}$. Now, for $\Delta=0$, we have the isotropic ($XXX$) spin exchange interaction $(c_x=c_y=c_z)$ whose eigenenergy spectrum is gapless \cite{Samaj2013}. So the first ``excited state'' with nearest eigenvalue to 1 goes as $1-c_1/L^2$ where $c_1$ is a constant (one $x$-polarized magnon excitation with the smallest pseudomomentum $k=2\pi/L$). Therefore, we readily identify $\beta=2$ for the $L$-dependence of $\lambda_1$ in the presence of $U(1)$ symmetry in our model when $\Delta=0$. The Thouless time here depends quadratically on the length of the lattice $t^*\simeq L^2/c_1$.

For $\Delta \ne 0$, the mapping in Eq.~(\ref{map}) in the Trotter regime consists of an anisotropic Heisenberg model with anisotropy $\eta = (J^2 + \Delta^2)/(J^2 - \Delta^2)$, $|\eta|>1$, which has a finite and system-size independent gap in the energy spectrum between the ground and first excited state \cite{Samaj2013}. Therefore, $\beta=0$ in the absence of $U(1)$ symmetry, and we have a finite ($L$ independent) Thouless time for our particle-number non-conserving model.

All this analysis for the $L$-dependence so far applies only to Trotter regime (of small $J,\Delta$ or in the continuous-time limit). Next, we investigate numerically the $L$-dependence of the eigenvalues of $\mathcal{M}$ for arbitrary $J,\Delta$. For this, we write  the elements of $\mathcal{M}$ as
\bea
\mathcal{M}_{\underline{n},\underline{n}'}= \left|\sum_{p}e^{-iE_p}\la \underline{n}|E_p\ra \la E_p|\underline{n}'\ra\right|^2, 
\eea
where $E_p$ and $|E_p\ra$ are respectively eigenvalues and eigenstates of $\hat{H}_1$ satisfying: $\hat{H}_1|E_p\ra=E_p|E_p\ra$. 

\begin{table}[h!]
  \begin{center}
    \begin{tabular}{c|c|c|c|c|c|c|c}
      \multicolumn{4}{c|}{$J=1,\Delta=0$} & \multicolumn{4}{c}{$J=1,\Delta=0.3$}\\ 
      \hline
      $L$ & $\lambda_1$ & $\lambda_2$ & $\lambda_3$ & $L$ & $\lambda_1$ & $\lambda_2$ & $\lambda_3$\\
      \hline
      10 & 0.653  & 0.5023 & 0.4275 &10 & 0.7594  & 0.6139 & 0.5288\\ 
      12 & 0.7495 & 0.6087 & 0.5263 &12 & 0.75938  & 0.6079 & 0.5677\\
      14 & 0.8115 & 0.6892 & 0.6098 &14 & 0.75938  & 0.6152 & 0.6041\\
      16 & 0.8535 & 0.7493 & 0.6939 &15 & 0.75938  & 0.6328 & 0.6026\\
    \end{tabular}
     \caption{Three largest eigenvalues $\lambda_1,\lambda_2,\lambda_3$ (excluding $\lambda_0=1$) of $\mathcal{M}$ for different lengths $L$ in the presence ($\Delta=0$) and absence ($\Delta \ne 0$) of $U(1)$ symmetry in our lattice model.}
    \label{table1}
  \end{center}
\end{table}

In Tab.~\ref{table1}, we list the first few eigenvalues of $\mathcal{M}$ (excluding the largest eigenvalue 1) for different system sizes in the presence and absence of $U(1)$ symmetry.
From the limited system sizes accessible in exact diagonalization, we find, when $J=1,\Delta=0$, that $\beta=1.86$ (using the largest three system sizes $L=12,14,16$) if we are fitting to 
$\lambda_1=1-1/t^*$ and $\beta=2.08$ if fitting to $\lambda_1=e^{-1/t^*}$. When $J=1,\Delta=0.3$, we find consistently $\beta=0.0$. We also observe in our numerics that $\beta \to 2$ as $J \to 0$ for the $U(1)$ symmetric model ($\Delta=0$) as predicted above from the analysis in the Trotter regime. Thus, we conclude that $\beta=2$ or $\beta=0$, respectively, in the presence or absence of $U(1)$ symmetry in our model for arbitrary $J,\Delta$. We further notice in Tab.~\ref{table1} that the other eigenvalues (e.g., $\lambda_2,\lambda_3$) are mostly $L$-independent for $J=1,\Delta=0.3$. For $\Delta=0$, we find that the numerical values of $\lambda_j$ rapidly fall with increasing $j$, and the $L$-dependence of $\lambda_j$ for small $j$s are somewhat similar to that of $\lambda_1$. The above results are for the Hamiltonian $\hat{H}_1$ with periodic boundary condition on real space. When we take open boundary condition, we find $\lambda_1=0.9042, 0.9329, 0.9504, 0.9619$ respectively for $L=10, 12, 14, 16$ for $J=1,\Delta=0$, and $\lambda_1=0.7746, 0.7724, 0.7709, 0.7703$ respectively for $L=10, 12, 14, 15$ for $J=1,\Delta=0.3$. These values give, depending on the respective fit model $\beta=1.96$ (for $\lambda_1 = 1-1/t^*$) or $\beta=2.03$ (for $\lambda_1=e^{-1/t^*}$) for $J=1,\Delta=0$, and $\beta=0.0$ for $J=1,\Delta=0.3$. From all the above numerical analysis, we conclude that the Thouless time $t^*$ grows as $L^2$ in $U(1)$ symmetric models, and $t^*$ is $L$-independent in  $U(1)$ symmetry-broken models. 

Numerical investigations suggest that the stochastic many-body Markov chain $\mathcal{M}$ (\ref{markov}) is integrable not only for infinitesimal but also finite $J$ and $\Delta$. Specifically, 
level spacing statistics of spectra of ${\cal M}$ at fixed quasimomentum, for both cases shown in Tab.~\ref{table1}, is found to be perfect Poissonian statistics. 
This hints to existence of a new integrable system related to six vertex model.

\begin{figure*}
\includegraphics[width=\linewidth]{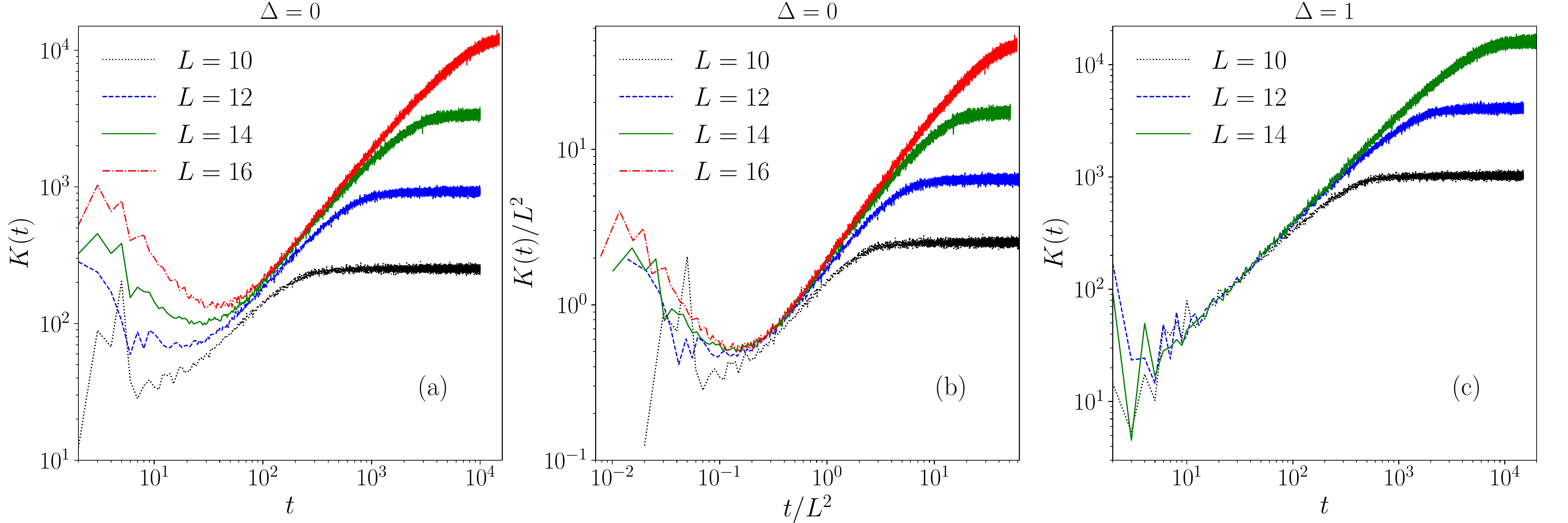}
\caption{Spectral form factor $K(t)$ for different system sizes $L$ of the kicked spinless fermion chain with $(\Delta=0)$ (a,b) and without $(\Delta=1)$ (c) particle-number conservation. Here, $J=1,U_0=15,\alpha=1.5$, while $\epsilon_i$ are random, normal distributed with zero mean and standard deviation $\Delta \epsilon=0.3$. We use open boundary conditions in real space, and take half-filling $N/L=1/2$ for $\Delta=0$. An averaging over $10^3$ realizations of disorder is performed. In (b) we show data collapse in scaled time $t/L^2$.}
\label{Fig1}
\end{figure*}

To validate our main analytical result (\ref{Mmat},\ref{series}), we finally perform exact numerical computations of $K(t)$ in our model~(\ref{ham1}) in the presence and absence of $U(1)$ symmetry. We calculate $K(t)$ numerically using Eq.~(\ref{SFF}). For that, we evaluate the matrix representation of $\hat{U}$ (\ref{EvU}) as:
\bea
\hat{U}_{\underline{n},\underline{n'}} = \sum_{p}\la\underline{n}|E_p\ra \la E_p|\underline{n}'\ra e^{-iE_p-i\theta_{\underline{n}'}}.
\eea
We take the $t$-fold matrix product and compute the trace of $\hat{U}^t$ for different $t$. We then repeat the evaluation of $|{\rm tr}\hat{U}^t|^2$ for different realizations of $\epsilon_i$, and finally take average over the disorder realizations. The numerically calculated $K(t)$ for different $L$ are plotted in Fig.~\ref{Fig1} for $\Delta=0$ and $\Delta=1$. Though we take here a half-filling for $\Delta=0$ case, our results are insensitive to the exact value of fixed filling fraction $N/L$ for large $L$ as long as we are sufficiently far away from either empty of full band.

For $\Delta=1$, we observe $K(t)$ to grow linearly with time before saturating around $t_H=\mathcal{N}=2^L$. The temporal growth of $K(t)$ for $\Delta=1$ in Fig.~\ref{Fig1}(c) at $t\ll t_H$ is independent of $L$ which confirms our  above analytical prediction based on the RPA. The temporal growth of $K(t)$ for $\Delta=0$ in Fig.~\ref{Fig1}(a) at $t \ll t_H$ depends strongly on $L$, and $K(t)$ again grows in time before saturating around $t_H=\mathcal{N}={L\choose L/2}$ for the half-filled case considered here. To investigate the $L$-dependence of the initial temporal growth of the computed $K(t)$, we plot $K(t)/L^2$ with $t/L^2$ in Fig.~\ref{Fig1}(b). We observe a nice data collapse for various $L$ and $t < t_H$ which confirms our above predicted $L$-dependence of $K(t)$ 
for the particle-number conserving case. Therefore, we find the analysis using the RPA is in agreement with the exact numerical predictions. Nevertheless, we need to consider a large enough value of interaction strength $U_0$ (e.g., 15) and appropriate falloff exponent $\alpha$ (e.g., 1.5), and use averaging over a large number of disorder realizations to match numerical results with the theoretical prediction applying the RPA.  

It is apt here to compare our results with Ref.~\cite{FriedmanPRL2019}. Interestingly, Ref.~\cite{FriedmanPRL2019} derived the same scaling of the Thouless time with the system size ($t^* \propto L^2$) in the $U(1)$ symmetric model, even though there are significant differences between the two models as well as in the methods. For example, while they work with nearest-neighbor interactions and need local Hilbert space dimension $q \to \infty$ for the Haar averaging, we can use $q=2$ but we need long-range interactions and disorder averaging to apply RPA. Second, Ref.~\cite{FriedmanPRL2019} find that the effective model is the integrable Trotterized XXX spin-1/2 chain (i.e., six vertex model). At the same time, in our case, the mapping to the XXX chain works only in the limit $J \to 0$, while for finite $J$, it does not correspond to the same Trotterized XXX chain (though it still seems to correspond to a Bethe ansatz integrable model as suggested above).    

In conclusion, our study extends the recent effort to identify microscopic mechanisms of quantum chaos to a many-body fermionic lattice system with nearest-neighbor hopping processes and long-range pairwise interactions in the presence or absence of conserved particle number. A new dynamical chaos mechanism has been found which maps the SFF $K(t)$ to an average recurrence probability of a classical Markov chain with transition probabilities given as square-moduli of hopping amplitudes.
There can be many interesting further questions and generalizations of the present study: (i) One can systematically compute subleading contributions to $K(t)$, Eq. (\ref{Mmat}), beyond $t$ cyclic and $t$ anti-cyclic permutations generalizing diagramatics of Ref.~\cite{KosPRX2018}. (ii) It would be straightforward to generalize our study to bosonic chains, at least in the particle-number conserving case.  (iii) It would be exciting to investigate the effect of $U(1)$ symmetry on $K(t)$ and $t^*$ in a locally (nearest-neighbor) interacting many-body quantum system with finite local Hilbert space \cite{BertiniPRL2018}. (iv) Finally, long-range interacting systems with the disorder have been investigated in the recent years in the context of many-body localization transition \cite{BurinPRB2015,Singh2017}, and our results provide a useful tool to investigate the ergodic phase of such systems.
\\\\
We thank P. Kos and R. Singh for discussions. This work has been supported by the European Research Council under the Advanced Grant No. 694544 – OMNES, and by the Slovenian Research Agency (ARRS) under the Programme P1-0402. DR also acknowledges the funding from the Department of Science and Technology, India via the Ramanujan Fellowship. This work was supported in part by the International Centre for Theoretical Sciences (ICTS) during a visit for participating in the program - Thermalization, Many body localization and Hydrodynamics (Code: ICTS/hydrodynamics2019/11).

\bibliography{bibliographyRMT}
\end{document}